# Characteristic Scales, Scaling, and Geospatial Analysis

Yanguang Chen

(Department of Geography, College of Urban and Environmental Sciences, Peking University, 100871, Beijing, China. Email: chenyg@pku.edu.cn.)

**Abstract**: Geographical phenomena fall into two categories: scaleful phenomena and scale-free phenomena. The former bears characteristic scales, and the latter has no characteristic scale. The conventional quantitative and mathematical methods can only be effectively applied to scaleful geographical phenomena rather than the scale-free geographical phenomena. In this paper, a comparison between scaleful geographical systems and scale-free geographical systems are drawn by means of simple geographical mathematical models. The main viewpoints are as below. First, the scaleful phenomena can be researched by conventional mathematical methods, while the scale-free phenomena should be studied using the theory based on scaling such as fractal geometry; Second, the scaleful phenomena belong to distance-based geo-space, while the scale-free phenomena belong to dimension-based geo-space; Third, four approaches to distinguish scale-free phenomena from scaleful phenomena are presented, including scaling transform, probability distribution, autocorrelation and partial autocorrelation functions, and ht-index. In practice, a complex geographical system usually possesses scaleful aspects and scale-free aspects. Different methodologies must be adopted for different types of geographic systems or different aspects of the same geographic system.

# 1. Introduction

Geographical phenomena can be divided into two categories: one is those bearing characteristic scales, and the other is those having no characteristic scales. The former may be termed *scaleful phenomena*, while the latter is called *scale-free phenomena*. To research a geographical phenomenon, we should describe it, and then, try to understand it. Scientific research is no more than two links, one is description, the other is understanding (Gordon, 2005; Henry, 2002). If we really understand the phenomenon, we can explain its behavior or predict its development, and thus new knowledge appears. A scaleful phenomenon can be characterized by conventional measures such as length, area, volume, eigenvalues, average values, and standard deviations. Further, we can model it using conventional mathematical methods and attempt to explain the causality behind the phenomenon. Unfortunately, for the scale-free phenomena, the conventional measures and mathematical methods usually fail. The essence of scale-free phenomena is the property of scale dependence. In other words, changing measurement scales will lead to different measurement results. Uncertain measurements lead to uncertain descriptions, which in turn lead to uncertain understanding. Consequently, the interpretation about it is not accurate and the prediction for it is not reliable. During the quantitative revolution of geography (1953-1976), remarkable achievements were made for the development of discipline. Unfortunately, many mistakes also appeared. One of mistakes is that the scaleful phenomena were confused with the scale-free phenomena. All geographical phenomena are treated as possessing characteristic scales and were quantitatively analyzed or mathematically modeled by means of conventional ways.

Nowadays, we can discriminate the scale-free geographical phenomena to a degree from the scaleful phenomena using scientific means. The scale-dependent problems should be solved by the ideas from scaling analysis. Based on a series of measurement scales and a series of corresponding measurement results, a scaling relation can often be constructed and one or more scaling exponents can be obtained. Using the scaling exponents to replace the characteristic scales, we can make efficient spatial analysis for geographical systems. The tools of scaling analysis include fractal geometry, allometry theory, complex network theory, wavelet transform, and renormalization group. Among all these tools for scaling analysis, the most important one is fractal geometry (Batty and Longley, 1994; Frankhauser, 1994). This is a kind of mature and systematic approach for scaling analysis (Chen, 2008a). How to integrate various mathematical theories to make scaling analyses of geographical systems? The precondition is to know the differences and similarity between scaleful phenomena and scale-free phenomena. In order to understand the nonlinearity, we must first understand the linearity; similarly, to understand the scaling idea, we must first understand the characteristic scale concept. This paper is devoting to clarifying the characteristic scales and scaling in geographical research. The aim is at how to make effective geographical spatial analysis. In Section 2, a comparison between characteristic scale and scaling is drawn, and three basic approaches to distinguishing scale-free phenomena from scaleful phenomena is proposed. In Section 3, the analytical processes defined in distance-based geo-space and the analytical processes defined in dimension-based geo-space are illuminated. In Section 4, several related questions are discussed, and finally, the discussion is concluded by summarizing the main points of this work.

## 2. Mathematical modeling ideas

### 2.1 Characteristic scales

Conventional mathematical modeling and quantitative analysis are based on characteristic scales. If and only if a system bears characteristic scales, we can model it using usual mathematical methods and analyze it by statistical techniques based on observational data. A characteristic scale is always represented by a 1-dimensional measure; therefore, it is termed *characteristic length* in scientific research (Hao, 1986; Liu and Liu, 1993; Takayasu, 1990; Wang and Li, 1996). Characteristic length includes side length, radius, diameter, height, eigenvalue, average value, and standard deviation. For a geometric object, if we can find a determinate side length, major axis, equivalent radius, and so on, to represent its basic spatial measure, we will say that it has characteristic length; for a correlation matrix, if we can find determinate eigen values (characteristic roots) to represent its geometric characteristic quantities, we will say that it bears characteristic length; for a probability distribution, if we can find determinate mean and standard deviation to represent its statistical characteristic quantities, we will say that it possesses characteristic length. On the other hand, if we find a characteristic length for a geometric object, we can predict its total length, area, volume, or density; for a correlation matrix, if we find characteristic roots for it, we can predict the corresponding variable relationships; for a probability distribution, if we find effective average value and standard deviation for it, we can determine its probability structure and predict the pattern and process of a system's evolution (Table 1).

**Table 1 Characteristic lengths in calculus, linear algebra, and probability theory and statistics**

| Advanced mathematics | Characteristic length | Mathematical property | Typical case in geography |
|---|---|---|---|
| **Calculus** | Side length, radius, height, etc. | Length, area, volume, density can be measured determinately | Radius of equivalent circle of a lake is the characteristic radius of the lake |
| **Linear algebra** | Eigenvalues (characteristic roots) | Eigenvalues and eigenvectors can be found determinately | Moran's *I* for spatial autocorrelation is an eigenvalue of spatial correlation matrix |
| **Probability theory & statistics** | Average value, standard deviation | Average values and standard deviations can be calculated determinately | The reciprocal of density decay rate of Clark's model is an average value |

In the past, scientists believed that all natural phenomena had characteristic scales. Socioeconomic phenomena were regarded as adhering to this default assumption. Therefore, the distinction between "scaleful phenomena" and "scale-free phenomena" is superfluous. In this case, for a long time, geographers conduct mathematical modeling and quantitative analysis based on the assumption that all geographical phenomena have characteristic scales. The discovery of fractals changed this understanding and resulted in new epistemology. Scientists realize that many complex phenomena have no characteristic scale. Such phenomena are regarded as scale-free phenomena. "Scale-free" concept emerges as the times require. The concept suggests that the phenomena that can be modeled by conventional mathematical methods and analyzed by traditional quantitative technologies are scale-based phenomena. For the convenience of expression, we need a concept opposite to the term "scale free". Just because of this, I coined a word "scaleful" to mean that a thing has one or more characteristic scales. The number of characteristic scales depends on the number of independent coordinates needed to describe a geographical phenomenon. For geographical lines or isotropic geographical surfaces, we need a characteristic scale (e.g., side length, radius); for anisotropic geographical surfaces, we need two characteristic scales (e.g., base length and height); for *n*-dimensional linear geographical systems, we need *n* characteristic scales (*n* eigenvalues) in principle; for spatial probability distributions, we usually need two to three or more characteristic scales (mean value and standard deviation as well as covariance).

## 2.2 Scaling and symmetry

The essence of scaling is a form of symmetry, that is, invariance under contraction or dilation transformation. This symmetry is termed scaling symmetry, indicating scale invariance (Batty and Longley, 1994; Chen, 2008a; Mandelbrot, 1982). Typical scaling phenomena are fractals, and self-similarity is just a kind of invariance under contraction or dilation (Mandelbrot, 1989). If a system bears some type of scaling property, then it can be modeled by a function, which conforms to the following relation

$$\mathbf{T}f(x) = f(\gamma x) = \gamma^a f(x) = \lambda f(x), \quad (1)$$

where $f(x)$ denotes the function, $x$ refers to an argument, $\mathbf{T}$ represents a scaling transform, $\gamma$ is a scale factor for the scaling transform, $a$ is a scaling exponent, and $\lambda = \gamma^a$ is the eigenvalue of the transform. A scaling transform is actually a contraction-dilation transform, and an eigenvalue is also called characteristic value. The most basic and common solution to the functional equation, i.e., equation (1), is power functions. For example, for the Zipf's law of rank-size distribution, we have

$$S(k) = S_1 k^{-q}, \quad (2)$$

in which $k$ refers to rank, $S(k)$ denotes the corresponding size, $S_1$ is the proportionality coefficient indicating the maximum size, and $q$ is the Zipf exponent, namely, the scaling exponent of rank-size distribution. Applying a scaling transform to Zipf's law shows scale invariance as follows

$$\mathbf{T}S(k) = S(\gamma k) = S_1 (\gamma k)^{-q} = b^{-q} S_1 k^{-q} = \gamma^{-q} S(k) = \lambda S(k). \quad (3)$$

Apparently, Zipf's law follows the scaling law, and the eigenvalue is the function of Zipf's exponent, that is, $\lambda = \gamma^{-q}$. This suggests that the key parameter of Zipf's distribution is the scaling exponent rather than the proportionality coefficient. If we apply the scaling transform to other power laws, e.g., Hack's law, allometric growth law, Pareto distribution, gravity model based on power-law decay, we will find the similar results. Just because of this, power laws are treated as being equivalent to scaling in some literature. As a matter of fact, scaling is not limited to power laws.

However, many mathematical models in geography do not comply with the scaling relation. If we apply the scaling transform to the functions bearing characteristic scale parameters, the scale invariance will be invalid. For example, Gaussian function, Clark's law, gamma distribution function, logarithm-normal distribution function, and all that, do not take on invariance under the scaling transform. All these mathematical laws and distributions belong to scaleful category, which can be dealt with by quantitative techniques and modeling methods based on traditional higher mathematics. Among these functions, Gaussian distribution is the simplest distribution with characteristic scales. Scaleful systems are simple systems, which do not require scaling analysis. Geographical modeling and analyses based on characteristic scales and those based on scaling are quite distinct from each other in theory.

**2.3 Distinction between scaleful and scale-free phenomena**

How to judge whether a geographical phenomenon has characteristic scale? The first approach is to examine the scale dependence of spatial measurement. If the measured result is independent of measurement scale, element granularity, image resolution, sample size, and so on, the geographical phenomenon bears characteristic scale. In contrast, if the measured result depends significantly on measurement scales, etc., the geographical phenomenon has no characteristic scale. The second approach is to examine the probability distributions of geographical events. Generally speaking, the probability distributions of scaleful systems take on unimodal curves. Exponential distribution is an exceptional case, but an exponential decay curve defined in a 1-dimensional space can be converted into a unimodal distribution curve defined in a 2-dimensional space. In other words, a unimodal curve can be derived from the one-side exponential attenuation curve. On the contrary, for a scale-free system, the probability distribution is always a long-tailed curve. The third approach is to examine autocorrelation function (ACF) and partial autocorrelation function (PACF). For a scaleful phenomenon, the ACF is a trailing-out curve, while the PACF is a cut-off curve; for a scale-free phenomena, both the ACF and PACF are trailing-out curves. The fourth approach is to compute the head/tail index (ht-index). The ht-index was proposed and developed by Jiang and his co-workers ([Jiang and Yin, 2014](#)). This measurement can be employed to quantitatively discriminate scale-free geographical phenomena from scaleful geographical phenomena. If the ht-index value is less than 3, it is considered to be scaleful phenomenon; and in contrast, if the ht-index value is greater than 3, it is considered to be scale-free phenomenon.

Two typical models can be employed to explain characteristic scales and scaling in geographical spatial analysis. One is Clark's model based on negative exponential decay ([Clark, 1951](#)), and the other is Smeed's model based on inverse power law decay ([Smeed, 1963](#)). The two models can be employed to describe urban density ([Batty and Longley, 1994](#)). Based on digital maps, urban population density follows Clark's law, which can be expressed a negative exponential function ([Clark, 1951](#)). Clark's law can be expressed as below:

$$\rho(r) = \rho_0 e^{-r/r_0}, \tag{4}$$

where $\rho(r)$ denotes the population density at the distance $r$ from the center of city ($r=0$). As for the parameters, $\rho_0$ refers to the urban central density $\rho(0)$, and $r_0$ to the characteristic radius of the population distribution (Chen, 2015a; Takayasu, 1992). In contrast, based on digital maps or remote sensing images, traffic network density satisfies Smeed's model, which can be expressed an inverse power law (Smeed, 1963). Smeed's model can be expressed as follows

$$\rho(r) = \rho_1 r^{D-d} = \rho_1 r^{-\alpha}, \tag{5}$$

where $\rho(r)$ denotes the traffic network density at the distance $r$ from the center of city ($r=0$), and $d=2$ is the Euclidean dimension of embedding space. As for the parameters, $\rho_1$ refers to traffic network density near city center, $D$ to the fractal dimension of traffic network, and $\alpha=d-D$ to the scaling exponent of the urban traffic network (Chen $et\ al$, 2019). Both the two models can give one-side attenuation curves. However, Smeed's inverse power-law decay is essentially different from Clark's negative exponential decay (Figure 1).

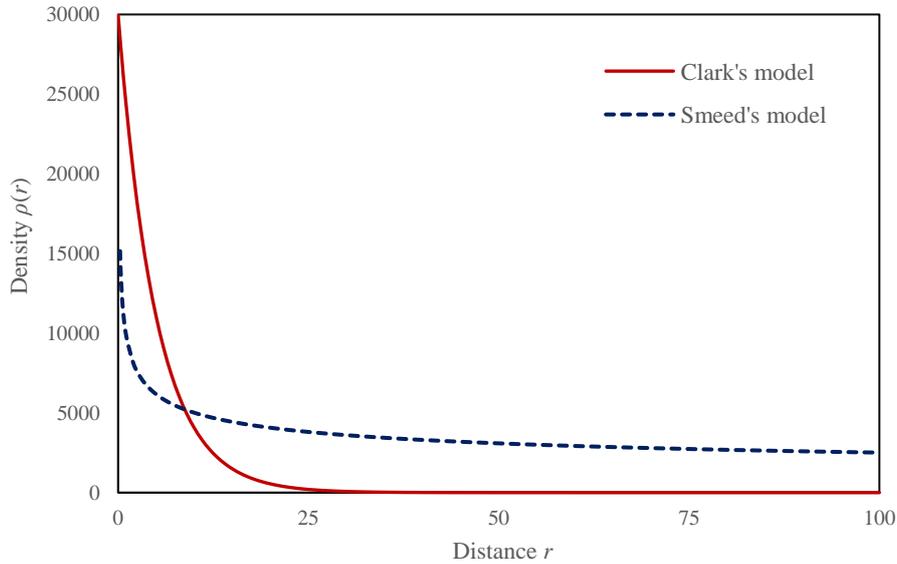

**Figure 1 A comparison between Clark's negative exponential decay and Smeed's inverse power-law decay**

**Note**: (1) For Clark's model, the parameter values are as follows: $\rho_0=30000$, $r_0=5$. (2) For Smeed's model, the parameter values are as follows: $\rho_1=10000$, $D=1.7$, $\alpha=2-D=0.3$.

The differences between negative exponential functions and inverse power laws are as follows. First, inverse power laws comply with scaling law, while negative exponential functions do not conform to scaling law. Applying scaling transform to Clark's model yields

$$\mathbf{T}\rho(r) = \rho(\gamma r) = \rho_0 e^{-\gamma r/r_0} = \rho_0^{1+\gamma}(\rho(r))^{\gamma} \neq \lambda\rho(r), \tag{6}$$

where $\mathbf{T}$ denotes scaling transform operator, $\gamma$ represents a scale factor. We cannot derive a scaling result $\lambda\rho(r)$, in which $\lambda$ refers to the eigenvalue under scaling transform $\mathbf{T}$. In contrast, applying scaling transform to Smeed's model yields

$$\mathbf{T}\rho(r) = \rho_1(\gamma r)^{D-d} = \gamma^{D-d}\rho_1 r^{D-d} = \gamma^{D-d}\rho(r) = \lambda\rho(r), \tag{7}$$

where $\lambda=\gamma^{D-d}$ denotes the eigenvalue under scaling transform $\mathbf{T}$. This implies that Smeed's model satisfies scaling relation, equation (1). On the contrary, applying translation transform to Clark's model yields

$$\mathbf{T}^*\rho(r) = \rho(r+\varsigma) = \rho_0 e^{-(r+\varsigma)/r_0} = e^{-\varsigma/r_0}\rho_0 e^{-r/r_0} = \lambda\rho(r), \tag{8}$$

where $\mathbf{T}^*$ denotes translation transform operator, $\zeta$ represents a translation factor, and $\lambda=\exp(-\zeta/r_0)$ is the eigenvalue under translation transform $\mathbf{T}^*$. A conclusion can be reached that Clark's model follows translational symmetry law, while Smeed's model follows scaling symmetry law.

Second, a unimodal curve indicating characteristic scale can be derived from the exponential attenuation curve. However, we cannot derive a unimodal curve from power-law attenuation curve. Let's examine Clark's model and derive a characteristic length by mathematical transformation. The procedure is as follows. **Step 1**: define equivalent circles. By the idea from statistical average, the area of equivalent circles can be defined as follows

$$A(r) = \pi r^2, \tag{9}$$

where $A(r)$ denotes the area within the radius $r$, and $\pi$ is the circular constant. **Step 2**: determine cumulative population. Finding the areal integral of $\rho(r)$ over $r$ based on equation (4) yields

$$P(r) = 2\pi\int_0^r x\rho(x)\mathrm{d}x = 2\pi\rho_0\int_0^r xe^{-x/r_0}\mathrm{d}x = 2\pi r_0^2\rho_0[1-(1+\frac{r}{r_0})e^{-r/r_0}], \tag{10}$$

where $P(r)$ refers to the total population within the equivalent circle with radius $r$. If $r$ is large enough, we will have $P(r)=2\pi\rho_0 r_0^2$. This suggests that urban total population can be predicted by the characteristic radius, $r_0$. **Step 3**: derive linear density. Differentiating $P(r)$ with respect to $r$ yields

$$\frac{\mathrm{d}P(r)}{\mathrm{d}r} = 2\pi\rho_0 re^{-r/r_0} = 2\pi r\rho(r) = \rho(r)\frac{\mathrm{d}A(r)}{\mathrm{d}r}, \tag{11}$$

which gives the circular density of urban population distribution. This is s special gamma function.

Clark's model reflects point density $\rho(r)$, while equation (11) give linear density $dP(r)/dr$. **Step 4**: search for the point of the extreme value. Taking the second derivative of equation (10) yields

$$\frac{d^2 P(r)}{d r^2} = 2\pi\rho_0 e^{-r/r_0}(1-\frac{r}{r_0}) = 2\pi\rho(r)(1-\frac{r}{r_0}) \quad . \tag{12}$$

If $d^2P(r)/dr^2=0$, then we have $r=r_0$. This suggests that $dP(r)/dr$ reaches the maximum value at radius $r_0$. **Step 5**: draw density curves. The peak of the curve gives the characteristic value. In fact, the point density curve is a one-side monotonic attenuation curve, while the linear density, i.e., circular density, is a unimodal curve, with a peak value at $r=r_0$ (Figure 2). Further, the average density distribution function based on Clark's law is as below

$$\rho_a(r) = \frac{P(r)}{A(r)} = \frac{2r_0^2 \rho_0}{r^2}[1-(1+\frac{r}{r_0})e^{-r/r_0}] \to 2\rho_0(\frac{r_0}{r})^2, \tag{13}$$

This indicates that the average population density takes on a power function, which is independent of fractals. Fractals is associated with power laws, but power laws does not necessarily imply fractals and fractal dimension.

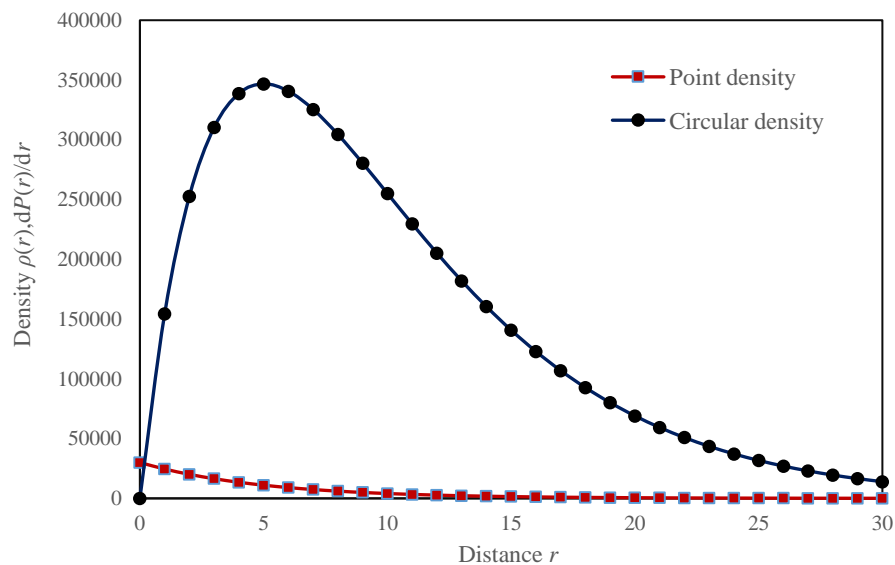

**Figure 2 The circular density curve based on Clark's negative exponential model**

**Note**: In Clark's model, the parameter values are taken as follows: $\rho_0$=30000, $r_0$=5.

In contrast, applying the above-shown procedure to Smeed's model cannot gives a characteristic value. By integral, the cumulative population is

$$P(r) = 2\pi \int_0^r x\rho(x)\mathrm{d}x = 2\pi\rho_1 \int_0^r x^{D-d+1}\mathrm{d}x = \frac{2\pi\rho_1}{D} r^D. \tag{14}$$

Under these circumstances, the total population of a city cannot be predicted by a typical radius. Taking the first derivative of equation (13) yield linear density as below

$$\frac{\mathrm{d}P(r)}{\mathrm{d}r} = 2\pi\rho_1 r^{D-1} = 2\pi r \rho(r) = 2\pi r \frac{\mathrm{d}A(r)}{\mathrm{d}r}. \tag{15}$$

Taking the second derivative of equation (13) yields

$$\frac{\mathrm{d}^2 P(r)}{\mathrm{d}r^2} = 2\pi(D-1)\rho_1 r^{D-2} = 2\pi(D-1)\rho(r) > 0. \tag{16}$$

According to the meanings of variable and parameters, $\mathrm{d}^2P(r)/\mathrm{d}r^2 \neq 0$. This suggest that no extreme value can be found, and the linear density curve is a monotonic increasing curve (Figure 3). Moreover, the average density distribution function based on Smeed's model is as below

$$\rho_a(r) = \frac{P(r)}{A(r)} = \frac{2\pi\rho_1}{D\pi r^2} r^D = \frac{2\rho_1}{D} r^{D-2}, \tag{17}$$

which suggests that

$$\rho_a(r) \propto \rho(r) \propto \frac{\mathrm{d}^2 P(r)}{\mathrm{d}r^2}. \tag{18}$$

No characteristic radius can be found from equation (17). The average network density distribution can give fractal dimension of traffic networks.

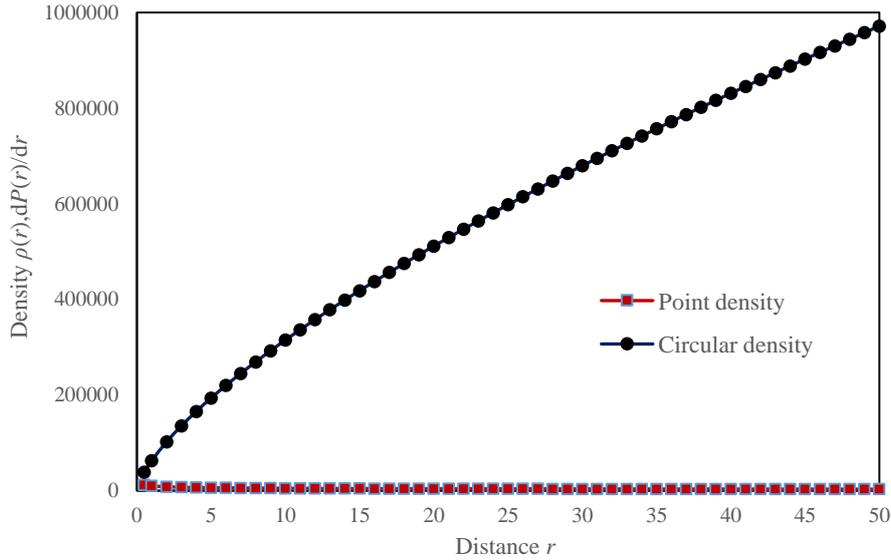

**Figure 3 The circular density curve based on Smeed's inverse power-law model**

Note: In Smeed's model, the parameter values are taken as follows: $\rho_1$=10000, $D$=1.7, $\alpha$=2-$D$=0.3.

Third, a comparison can be drawn between ACF and PACF of a data series indicative of distance decay. The differences between the power-law distance decay and exponential distance decay can be brought to light by using ACF and PACF of time series or spatial series. ACF includes both direct autocorrelation and indirect autocorrelation, while PACF only indicate direct autocorrelation, reflecting no indirect autocorrelation in theory (Chen, 2008b; Chen, 2015b; Diebold, 2007). For the exponential decay series, the ACF displays gradual one-sided damping, but the PACF cuts off at a displacement of 1 (Chen, 2008b; Chen, 2012a; Chen, 2015b). The former is a trailing-out curve while the latter is a cut-off curve (Figures 4). This suggests that the direct spatial action of a location based on Clark's model is significantly localized and cannot reach distant locations. This is inconsistent with the law of geography put forward by Tobler (1970, 2004). In contrast, the curves of ACF and PACF of the power-law decay series tail off gradually. Both the two curves are trailing-out curves (Figures 5). Differing from Clark's model, Smeed's model is follows the first law of geography, indicating action at a distance. Generally speaking, the exponential function indicates a simple system with characteristic lengths, while the power law implies a complex system without characteristic length (Barabási, 2002; Barabasi and Bonabeau, 2003; Chen, 2015a; Chen, 2015b).

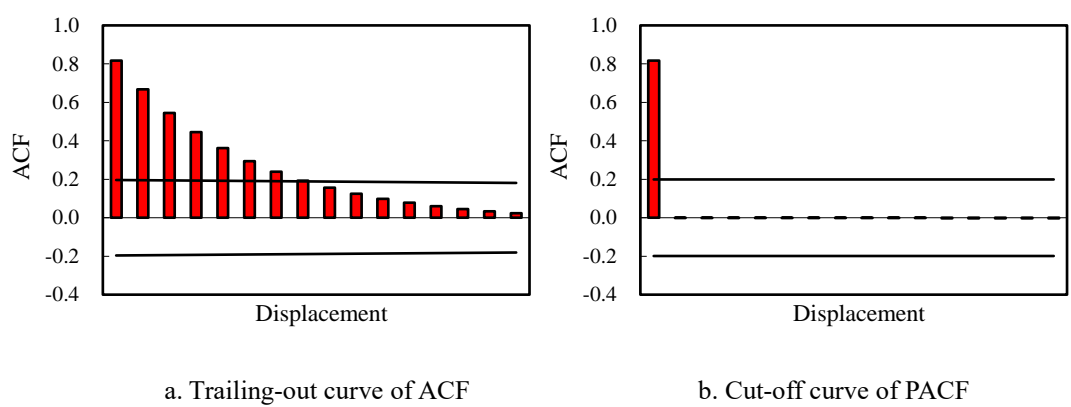

a. Trailing-out curve of ACF　　　　b. Cut-off curve of PACF

**Figure 4 The schematic histograms of spatial ACF and PACF based on Clark's negative exponential model**

Note: (1) In Clark's model, the parameter values are taken as follows: $\rho_0$=30000, $r_0$=5. (2) ACF and PACF are dimensionless quantities varying -1 to 1. The displacement unit depends on the distance unit such as kilometer. (3) The two horizontal lines represent the "double-standard-error bands (DSEB)" of ACF or PACF. If the ACF or PACF comes between the two DSEB lines, the ACF or PACF can be omitted at the significance level $\alpha$=0.05.

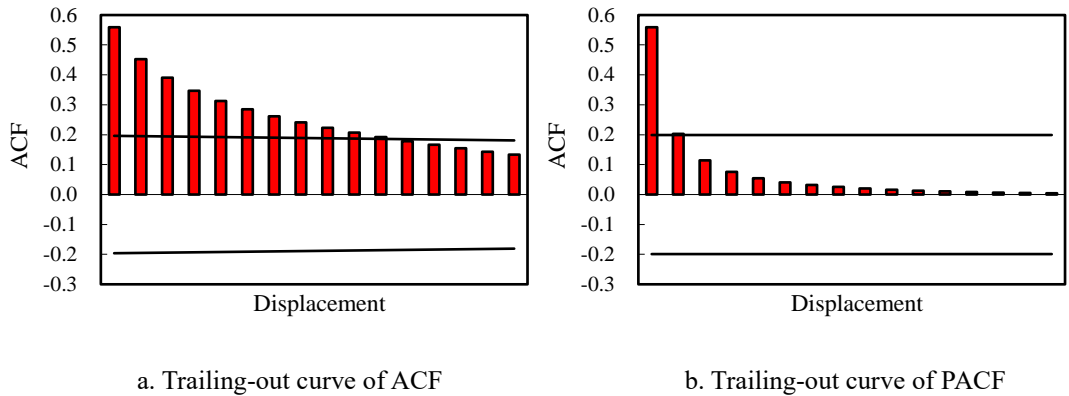

a. Trailing-out curve of ACF    b. Trailing-out curve of PACF

**Figure 5 The schematic histograms of spatial ACF and PACF based on Smeed's inverse power function**

**Note**: (1) **Note**: In Smeed's model, the parameter values are taken as follows: $\rho_1$=10000, $D$=1.7, $\alpha$=2-$D$=0.3. The minimum distance is set as $r_0$=0.1, and the corresponding density is $\rho_0$= 19952.6231. (2) The explanation about symbols is the same as those in Figure 4.

## 3. Two types of spatial analysis

### 3.1 Distance-based geo-spatial analysis

Geographical research depends heavily on spatial analysis, which in turn depends heavily on spatial quantification and spatial modeling. However, spatial modeling is one of the most difficult problems of mathematical modeling in scientific research (Waldrop, 1992). In fact, there are three difficulties in mathematical modeling, that is, spatial dimension, time lag, and interaction of multiple elements. Spatial dimension implies network structure, time lag implies response delay, and interaction implies element coupling. These three aspects of problems result in nonlinearity and thus result in complexity. Traditional geographical spatial analysis relies heavily on distance variable, and thus the conventional geographical space in theory can be regarded as distance-based space (Johnston, 2003). Central place models, gravity models, spatial interaction analysis, spatial autocorrelation analysis, and all that, are based on distance measurement. In this case, the distance is defaulted to be a significant spatial quantity with characteristic scales.

For the scaleful geographical systems, geographical space can be converted into mathematical space based on distance. If geographical distance can be measured efficiently, we find a characteristic value based on distance variable and make spatial analysis. For example, the well-known Moran's index for spatial autocorrelation analysis is just an eigenvalue of generalized spatial correlation matrix. Concretely speaking, we have

$$Mz = Iz, \qquad (19)$$

where $M^*=zz^\mathrm{T}W$ is generalized correlation matrix, $z$ is standardized size vector, and $I$ denotes Moran's index. The generalized correlation matrix is defined as below:

$$M = zz^\mathrm{T}W, \qquad (20)$$

where $W$ is the unitized spatial weight matrix based on distance matrix. According to equation (19), Moran's $I$ proved to be the eigenvalue of $M$, and the corresponding eigenvector is $z$ (Chen, 2013). If equation (20) is replaced by $M^*=z^\mathrm{T}zW =nW$, where $n$ is sample size or the number of spatial elements, then we will have $nWz=Iz$, Moran's $I$ will be decomposed into a set of eigenvalues indicating different directions: $\lambda_1, \lambda_2,\ldots, \lambda_n$. Using Moran's $I$ and the corresponding set of eigenvalues, we can make spatial autocorrelation analysis. The problem lies in that if the spatial distance matrix is uncertain, Moran's index will also be an uncertain quantity. In this instance, Moran's index, $I$, will lose the significance of eigenvalue.

In order to describe a geographical phenomenon and try to understand it, we had better model it using mathematical methods. The modeling results are so-called mathematical models. Models can be treated as the essence of science, and mathematical models represent the essence of models. In short, mathematical models are essential components in scientific research (Holland, 1998; Kac, 1969; Longley, 1999). A good mathematical model of a system is always involved with three scales: a macro scale indicative of environment, a micro scale indicative of interaction of elements, and the characteristic scale indicative of the inherent structure of the system (Hao, 1986). Correspondingly, the good model bears three parameters: environmental parameter, element parameter, and characteristic parameter (Figure 6). As an example, the logistic model of fractal dimension growth curve can be employed to illuminate the three types of parameters. For a growing fractal such as an urban agglomeration, the time series of fractal dimension values in different years forms a sigmoid curve (Chen, 2012b). The curve can be described by the logistic model as follows (Chen, 2018)

$$D(t) = \frac{D_{\max}}{1+(\frac{D_{\max}}{D_{(0)}}-1)e^{-\beta t}}, \qquad (21)$$

where $D(t)$ is the fractal dimension of time $t$, $D_{\max}$ is the capacity value of fractal dimension, i.e., the large value of fractal dimension, $D_{(0)}$ is the initial value of fractal dimension of time $t=0$, $\beta$ is the inherent growth rate. In this model, $D_{\max}$ represents the global parameter indicating the macro scale.

If the Euclidean dimension of the embedding space is $d=2$, then we have $D_{max} \leq 2$. The $D_{max}$ value depends on the environmental capacity. $D_{(0)}$ represents the local parameter indicating the micro scale. This parameter reflects the special property of a given growing fractal. The relative growth rate, $\beta$, is the characteristic parameter, indicating the typical scale of time. Equation (21) can be transformed into the following form

$$\frac{D_{max}}{D(t)} - 1 = (\frac{D_{max}}{D_{(0)}} - 1)e^{-\beta t}, \tag{22}$$

which indicates the ratio of residual dimension, $D_{max} - D(t)$, to the current dimension, $D(t)$. Equation (22) is a negative exponential function, and the $1/\beta$ represents the characteristic scale of time. For an exponential function, the relative growth/decay rate represents the characteristic parameter.

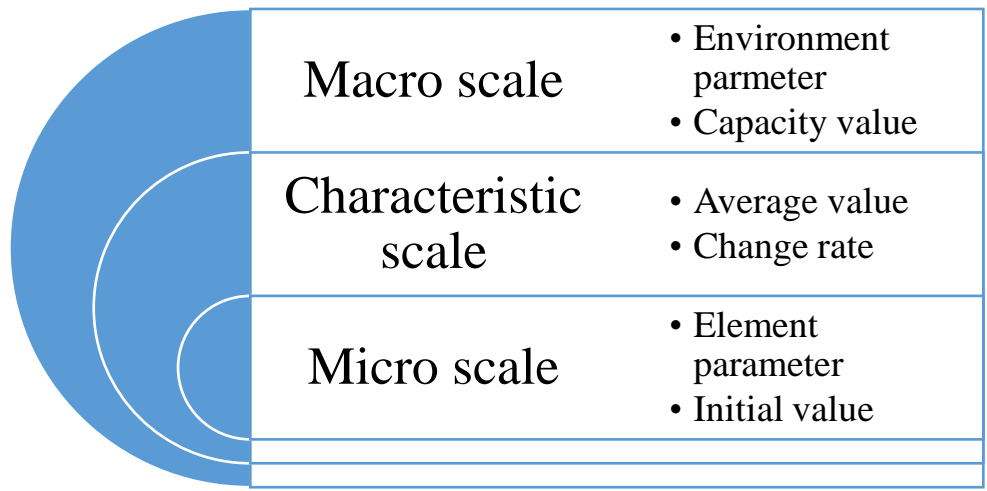

**Figure 6 Three scales for a good mathematical model of a geographical system**

**Note:** The parameter representing a characteristic scale is a key to analyzing a system and predicting its evolution.

### 3.2 Dimension-based geo-spatial analysis

For the geographical systems without characteristic scales, spatial analyses are often invalidly made on the base of distance variable. In this case, geographical space cannot be theoretically transformed into mathematical space based on distance. A solution to this problem is to replace the distance-based on space with dimension-based space. If the distance-based space is a kind of scaleful space, then the dimension-based space can be treated as a scale-free space. Clark's negative exponential distribution of urban population density can be described by a characteristic distance, $r_0$ (see equation (4)), but the Smeed's inverse power law distribution of urban traffic network density cannot be characterized by a distance (see equation (5)). As indicated above, we can utilize the characteristic radius to predict the total quantity of urban population, but we cannot use a typical radius to forecast the total length of urban traffic lines. The characteristic distance should be replaced by a scaling exponent, $\alpha=d-D$, which is equivalent to the fractal dimension $D=d-\alpha$ (Batty and Longley, 1994; Chen et al, 2019). In practice, if a geographical measurement depends on scales, or a geographical distribution follows power law, we meet with a scale-free geographical phenomenon. Power laws can be found in geographical fields everywhere. Differing from classic physics, geography dose not possess universal iron law. However, a number of models have a wide range of uses in geographical research and can be treated as mathematical "laws" in geography. Among these geographical laws, three ones are prominent, that is, distance-decay law (a geographical law for space), allometric growth law (a geographical law for time), and rank-size law (a geographical law for hierarchy). The three laws corresponds to three types of geographical space: real space (R-space), phase space (P-space), and order space (O-space) (Chen, 2008a; Chen, 2014; Chen et al, 2019). All these laws are mainly based on power laws, despite various variants in models.

Table 2 Three types of scientific laws in geography: space types and models

| Law | Orientation | Space | Model (example) |
| --- | --- | --- | --- |
| Distance-decay law | Space | Real space | Gravity models, spatial interaction models |
| Allometric growth law | Time | Phase space | Allometric scaling relations |
| Rank-size law | Hierarchy | Order space | Zipf's law, Pareto distribution |

**Note:** The scaling laws hidden behind central place models can be used to integrate these three geographical laws.

A set of typical power laws or mathematical models in geography based on power laws can be employed to illustrate the dimension-based spatial analysis. These power laws and models include allometric growth law, Zipf's law, gravity models based on distance decay law, and central place models. The allometric scaling relation can be expressed as follows

$$A=\eta S^{\sigma}, \tag{23}$$

where $A$ denotes area such as urbanized area, $S$ refers to size such as city population, $\eta$ is a proportionality coefficient, and $\sigma$ is scaling exponent. The scaling exponent is in fact the ratio of the fractal dimension of area measure $D_A$ to that of size measure $D_S$, that is

$$\sigma = \frac{D_A}{D_S}. \tag{24}$$

The allometric function can be used to model the scaling relationships between urban area and urban population (Batty, 2008; Lee, 1989; Nordbeck, 1971), the scaling relationship between island area and species number (Could, 1979; Wilson, 1992), the scaling relationship between the length of main river channel and the area of river basin (Feder, 1988; Hack, 1957; Mandelbrot, 1982), and so on. The hierarchies of central place networks can be modeled by the following power laws

$$N(r) = N_1 r^{-D}, \tag{25}$$

$$S(r) = S_1 r^{D_s}, \tag{26}$$

in which $r$ denotes the distance between two adjacent central places of the same level, $N(r)$ and $S(r)$ represents the number and size of central places, $N_1$ and $S_1$ are proportionality coefficients, and $D$ and $b$ are scaling exponents. In fact, $D$ is the fractal dimension of central place network, and $D_s$ is the average fractal dimension of central place sizes (Chen, 2015b). The common form of the gravity models is as below

$$I_{ij}(r) = K S_i S_j r^{-b}, \tag{27}$$

where $I_{ij}$ denotes attraction force, $S_i$ and $S_j$ represent size measures, $r$ refers to distance, $K$ is gravity coefficient, and $b$ is the distance decay exponent. The parameter relationships between Zipf's law, central place network, and gravity model are as follows

$$q = \frac{D_s}{D}, \tag{28}$$

$$b = qD = D_s, \tag{29}$$

in which $q$ is Zipf's exponent, the scaling exponent of rank-size distributions. Other symbols have the same meaning as stated above.

All these models are in fact classic models in geography. The commonalities of these models are as follows. First, these models are directly or indirectly based on distance variables. In the allometric growth model, the area measure contains the radius of equivalent circle, and the radius is actually a distance. Substituting Zipf's model, equation (2), into the allometric growth model, equation (23), we can reveal the spatial meaning of Zipf's distribution. Second, no characteristic distance can be determined. In other words, we cannot find a characteristic length in these models. This suggests that distance cannot serve as the characteristic quantity of geographical space. Third, power exponents represent fractal dimensions or ratios of one dimension to another dimension. Scaling exponents can be calculated by means of the power law relations between distance and the corresponding measures or between two different measures depending spatial sizes.

## 4. Questions and discussion

The scaleful geographical phenomena had been being confused with the scale-free geographical phenomena for a long time before emergence of fractal theory. Thus the dimension-based space was mistaken as distance-based space. The same mathematical tools were applied to geographical systems of different natures to make data processing, quantitative analysis, and mathematical modeling. Consequently, wrong description leads to wrong understanding, which in turn leads to wrong explanation and prediction. This paper tries to clarify the fact as below: if a geographical phenomenon possesses characteristic scales, we can analyze and model it using conventional mathematical tools in the traditional way; in contrast, if a geographical phenomenon has no characteristic scale, we should make analyses and models by means of the mathematical methods based on the idea of scaling (Table 3). Fractal geometry is the most effective tool to explore nonlinearity, singularity, and scale-free property. The basic nature of fractals is of no characteristic scales, or, scaling symmetry. A geographical system often takes on two opposite aspects of characters meantime: scaleful aspects and scale-free aspects. For example, a lake's boundary has no characteristic scale, but the area within the boundary bears characteristic scale. The boundary line of the lake can be modeled using Koch curve, a typical fractal line. However, the radius of the equivalent circle of a lake is the characteristic length of the lake's area. Urban form is more complex. Urban boundaries, urban land use patterns, urban traffic networks, and so on, bear no characteristic scale, but urban population density distribution may possess characteristic scales (Clark, 1951; Takayasu, 1990).

**Table 3 Mathematical methods of different natures and corresponding spheres of applications**

| Type | Space | Mathematical tool | Applied directions |
| --- | --- | --- | --- |
| **Methods** | Scaleful space: | Calculus | Geographical spatial patterns |

| based on characteristic scales | distance-based geo-space | Linear algebra | Geographical spatio-temporal processes |
| --- | --- | --- | --- |
| | | Probability theory & statistics | Geographical spatial and probability distributions |
| Methods based on scaling | Scale-free space: dimension-based geo-space | Fractal geometry | Geographical spatial patterns |
| | | Renormalization group | Geographical spatial patterns |
| | | Complexity network | Geographical networks |
| | | Allometry theory | Geographical evolution and hierarchy |
| | | Wavelet analysis | Geographical spatio-temporal processes |

As we known, western modern science begins with Galileo (1564-1642). So far, more than 300 years have passed. During this period, various mathematical equations and functions appeared in the scientific community. Among all kinds of mathematical expressions, three functions are the most representative ones, that is, Gaussian function, exponential function, and power function (Table 4). Before the 19th century, scientists focused on Gaussian function; in the 20th century, especially after the Second World War, scientists focused on exponential function; in the new century, scientists focused on power function, especially, the inverse power law in nature (Arbesman, 2012). Gaussian function indicates normal distribution, which is the mark of simplicity. A Gaussian function bears two parameters representing characteristic scales, that is, average value and standard deviation. In contrast, power function implies scale-free distribution, which is the mark of complexity (Goldenfeld and Kadanoff, 1999). No parameter represents characteristic scale, but the scaling exponent can be used to characterize complex systems. Exponential function comes between Gaussian function and power function. On the one hand, an exponential function bears parameter representing average value; on the other hand, a power function can be decomposed into a pair of exponential functions (Chen, 2014; Chen, 2015b). Due to the rise of complexity science, power function has become one of hot topics in scientific research in recent years. The behavior of complex systems always follow power laws, especially, the inverse power law of nature. Many scholars who do not understand the true meaning of power laws think that power laws is a fashionable topic in academia. For the rapid publication of papers, the scholars who misunderstand power laws have also written a lot of academic papers on power laws. Consequently, power laws were often over-identified, which led to over identification of scaling laws. In fact, a power law is always a solution to the scaling functional equation. One of the important reasons for over identification of power laws is that many scholars do not understand the mathematical principle of scaling. Another reason involves algorithm. The scaling exponent is usually estimated by the least squares method (LSM), and LSM is not helpful for researchers to distinguish power laws from fake power laws through scaling exponent values. Fortunately, a new algorithm based on maximum likelihood method (MLM) was developed by Clauset *et al* (2009) to evaluate scaling exponents. This method is useful to distinguish power laws. A study based on a series of allometric scaling relationships of Chinese cities from 1991 to 2014 shows the following results. If a power law relation is well developed, both LSM and MLM give similar scaling exponent values; in contrast, if the power law is less developed, the LSM will give approximate scaling exponent, while MLM give abnormal exponent value. This shows that LSM is suitable for approximate estimation of scaling exponents, while MLM is suitable for identifying true and false power-law relations (Table 5).

**Table 4 Three typical functions in modern and current scientific research**

| Stage | Period | Function | Characteristic scale | System nature |
|---|---|---|---|---|
| **First stage** | 19 century | Gaussian function | Average value and standard deviation | Simplicity |
| **Second stage** | 20 century | Exponential function | Average value | Coming between simplicity and complexity |
| **Third stage** | 21 century | Power function | Characteristic scale is replaced by scaling exponent | Complexity |

**Note:** Refer to Arbesman (2012).

**Table 5 A functional comparison between least squares method (LSM) and maximum likelihood method (MLM) for scaling research**

| Algorithm | Well-developed power law | Less-developed power law | Advantage | Disadvantage |
|---|---|---|---|---|
| **Least squares method (LSM)** | Valid scaling exponent (Reasonable results) | Approximate valid scaling exponent (Reasonable results) | Helpful for approximate estimation of scaling exponent | Not conducive to the identification of power laws |
| **Maximum likelihood** | Valid scaling exponent | Invalid scaling exponent | Helpful for identifying valid | Not conducive to the approximate |
| **method (MLM)** | (Reasonable results) | (Absurd results) | power laws | estimation of scaling exponent |

**Note:** The contents are summarized by referring to Chen and Feng (2017). The power laws in geographical systems are not iron law. They evolve from non-power law (less-developed scaling relation) to approximate power law and to real power law (well-developed scaling relation).

Complex geographical systems such as mountains, water systems, and cities bear no characteristic scale and follow power law. Power laws are basic relations for fractal parameter estimation. Fractal geometry can be employed to explore spatial complexity and find the association of simplicity with complexity (Batty and Longley, 1994; Frankhauser, 1994; Frankhauser, 1998). Wheeler (1983) once made a comment about fractal as below: "No one is considered scientifically literate today who does not know what a Gaussian distribution is or the meaning and scope of the concept of entropy. It is possible to believe that no one will be considered scientifically literate tomorrow who is not equally familiar with fractals." To my thinking, the last sentence should be improved as follows: "It is possible to believe that no one will be considered scientifically literate tomorrow who is not equally familiar with Pareto-Mandelbrot distributions and fractals." Pareto-Mandelbrot distribution is also termed Pareto-Zipf distribution (Frankhauser, 1998). Mandelbrot (1982) generalized the two-parameter Zipf model to a three-parameter Zipf model (Table 6). Today, Gaussian distribution and Pareto distribution represent two extreme and typical cases of geographical phenomena. The transition in the middle is exponential distribution. The traditional geographical thinking is mainly based on Gaussian distribution and Markov process (Ai, 1993; Chen, 1999; Jiang, 2015). In the past 30 years, geographical thinking has been quietly and gradually transformed (Jiang, 2015; Jiang, 2018). Maybe new theoretical paradigm will form in geography in the near future.

**Table 6 Two types of natural phenomena and corresponding information measures as well as spatial analyses**

| Type | Function | Information measure | System nature | Geographical analysis |
|---|---|---|---|---|
| **Scaleful phenomena** | Gaussian distribution | Entropy | Simplicity | Spatial analysis based on characteristic scales |
| **Scale-free phenomena** | Pareto-Mandelbrot | Fractals | Complexity | Spatial analysis based on scaling |

|  | distribution |  |  |  |  |

**Note:** Pareto-Mandelbrot distribution is also termed Pareto-Zipf distribution (Frankhauser, 1998), which is associated with fractal hierarchy. Refer to Wheeler (1983).

Geographical systems as complex spatial systems cannot be fully explained by means of the idea from reductionism. Where mathematical description is concerned, the reductionist methodology is always based on characteristic scales. A complex system bears at least three characters: irregular pattern, nonlinear process, scale-free distribution (scale dependence or no characteristic scale). These years, spatial complexity and scaling become important key words in geographical research. The main functions of scaling analyses are as follows. First, scaling analyses lay theoretical foundation of power laws (Feder, 1988; Liu and Liu, 1993; Mandelbrot, 1982); Second, scaling analysis can be used to find useful parameter solutions to some nonlinear equations (Chen, 2008; Takayasu, 1990); Third, scaling analyses can be employed to reveal scale-free features of complex spatial systems in empirical studies (Jiang and Yao, 2010). In recent years, the function and significance of scaling analysis have attracted more and more attention of geographers. A number of interesting findings, results, and viewpoints have been published (Arcaute *et al*, 2015; Bettencourt, 2013; Bettencourt *et al*, 2007; Bettencourt *et al*, 2010; Lobo *et al*, 2013; Jiang and Jia, 2011; Jiang and Liu, 2012; Louf and Barthelemy, 2014a; Louf and Barthelemy, 2014b; Pumain *et al*, 2006). Among various scaling analyses, the most frequently subject is allometric scaling relations in cities. The very important analytical tools is fractal theory. Some issues have reached a consensus, but others have different opinions. Anyway, the criteria for good scientific research are interest, novelty, and inspiration for insight. Scientists carry out a variety of explorations on geographical scaling. Many inspirational achievements came into being. In particular, Jiang and his co-workers tried to relax the definition of fractals and scaling, and fractal is redefined as below: *A set or pattern is fractal if the scaling of far more small things than large ones recurs multiple times* (Jiang and Yin, 2014). According to the new definition, the quantitative criterion of fractals is replaced by the head/tail index (ab. ht-index) (Jiang, 2013; Jiang, 2015): *the ht-index of a fractal set or fractal pattern is at least three* (Gao *et al*, 2017; Jiang and Yin, 2014). The new definition and criterion of fractals result in new understanding about scaling in geographical systems. In fact, due to spatial heterogeneity, it is impossible to describe complex processes and patterns of geographical systems using simple fractal models. Spatial homogeneity indicates spatial stationarity, that is, different parts in a geographical systems have the same probability structure comprising mean, standard deviation, and covariance. In contrast, spatial heterogeneity suggests spatial nonstationarity, i.e., different parts in the geographical system possess different probability structures: different means, different standard deviations, and different covariance. If the spatial heterogeneous system bear scaling properties, different parts will bear different scaling exponents. It is multifractal spectrums rather than simple fractal geometry can be employed to characterize spatial heterogeneity (Chen, 2008; Stanley and Meakin, 1988). The problem involve a number of academic concepts (Table 7). It is impossible to clarify all these questions in a few lines of words. This paper is devoted to discussing the connections and distinctions between scaleful phenomena and scale-free phenomena in geographical world. The shortcomings of this work lies in two aspects: one is that it is short of empirical analysis, and the other, this study is based on the conservational concepts of fractal and scaling. Owing to the limitation of space, the further theoretical analyses and positive studies remain to be made in future works.

**Table 7 The corresponding relationships between fractals, scaling, and spatial heterogeneity**

| Type | Subtype | Spatial distribution | Probability distribution | Scale property | Example |
|---|---|---|---|---|---|
| Homogeneity (Spatial stationarity) | Non-fractal (three typical nonfractal phenomena) | Uniform distribution | Absolute uniformity | Characteristic scale | Classical central place network |
| | | Random distribution | Statistic uniformity | Characteristic scale | Poisson distribution |
| | | Distance decay with characteristic length | Circular uniformity | Characteristic scale | Clark/Sherratt models of urban density |
| | Monofractal (unifractal) | Simple self-similar distribution | All fractal units are the same | Single scaling process | Fractal central place network |
| | | Distance decay based on scaling | Circular uniformity | Single scaling process | Smeed model of traffic network |
| Heterogeneity (Spatial nonstationarity) | Multifractals | Complex self-similar distribution | An fractal unit differs from another fractal unit | Multiple scaling process | Multifractal central place network |

# 5. Conclusions

Geospatial analysis is very interesting, but there are too many difficult problems to be solved before forming a theoretical framework for spatial modeling and quantitative analyses. The main points of this paper are as follows. **First, if a geographical phenomenon bears characteristic scales, it belongs to scaleful space and can be quantified and modeled using conventional mathematical methods.** An effective measure is associated with characteristic scales, and a good mathematical model and an effective quantitative analysis are often based on characteristic scales. A characteristic scale is always a 1-dimensional measure and termed *characteristic length*. The concrete characteristic lengths include radius of a circle, base length and height of a triangle, length and width of a rectangle, eigenvalues of correlation or distance matrix, and average value and standard deviation of a probability distribution. **Second, if a geographical phenomenon has no characteristic scale, it belongs to scale-free space and should be modeled and quantitatively analyzed by means of the methods based on scaling relations.** It is impossible to find the valid characteristic lengths for a scale-free phenomenon. The values of length, area, volume, density, eigenvalues, average values, and standard deviations of scale-free phenomena depend on the measurement scales. The basic properties of scale-free systems are scale-dependence, extreme probability distribution, and invariance under contraction or dilation transform. In this case, the characteristic length should be replaced by characteristic exponent, which is termed *scaling exponent*. **Third, the key to solving problems is to distinguish scaleful phenomena from scale-free phenomena in geographical systems.** Different types of mathematical methods have different spheres of application and should be used for different sorts of geographical phenomena. Geographical spatial analysis can be divided into two categories: distance-based geo-spatial analyses and dimension-based geo-spatial analysis. The former belongs to scaleful spatial analysis and the latter belongs to scale-free spatial analysis. The main approaches of identifying scale-free phenomena from various geographical phenomena include scaling transform, probability distribution, ACF and PACF analyses, and head/tail index (ht-index).

**Acknowledgements**

This research was sponsored by the National Natural Science Foundation of China (Grant No. 41671167). The support is gratefully acknowledged.

# References

Ai NS (1993). Mandelbrot landscape and Hurst phenomenon. In: Xin HW (ed.). *Fractal Theory and Its Application*. Hefei: University of Science and Technology of China Press [In Chinese]

Arbesman S (2012). *The Half-Life of Facts: Why Everything We Know Has An Expiration Date*. New York: Penguin Group

Arcaute E, Hatna E, Ferguson P, Youn H, Johansson A, Batty M (2015). Constructing cities, deconstructing scaling laws. *Journal of the Royal Society Interface*, 12(102): 20140745

Barabási A-L (2002). *Linked: The New Science of Network*. Cambridge: Persus Publishing

Barabasi A-L, Bonabeau E (2003). Scale-free networks. *Scientific American*, 288(5): 50-59

Batty (2008). The size, scale, and shape of cities. *Science*, 319: 769-771

Batty M, Longley PA (1994). *Fractal Cities: A Geometry of Form and Function*. London: Academic Press

Bettencourt LMA (2013). The origins of scaling in cities. *Science*, 340: 1438-1441

Bettencourt LMA, Lobo J, Helbing D, Kühnert C, West GB (2007). Growth, innovation, scaling, and the pace of life in cities. *PNAS*, 104(17): 7301-7306

Bettencourt LMA, Lobo J, Strumsky D, West GB (2010). Urban scaling and its deviations: Revealing the structure of wealth, innovation and crime across cities. *PLoS ONE*, 5(11): e13541

Chen YG (1999). Geography: Failure of the quantitative revolution and starting-up of the fractal studies. *Journal of Xinyang Teachers College (Natural Science Edition)*, 12(3): 310-314[In Chinese]

Chen YG (2008a). *Fractal Urban Systems: Scaling, Symmetry, and Spatial Complexity*. Beijing: Science Press (In Chinese)

Chen YG (2008b). A wave-spectrum analysis of urban population density: entropy, fractal, and spatial localization. *Discrete Dynamics in Nature and Society*, Volume 2008, Article ID 728420, 22 pages

Chen YG (2012a). On the four types of weight functions for spatial contiguity matrix. *Letters in Spatial and Resource Sciences*, 5(2): 65-72

Chen YG (2012b). Fractal dimension evolution and spatial replacement dynamics of urban growth. *Chaos, Solitons & Fractals*, 45 (2): 115-124

Chen YG (2013). New approaches for calculating Moran's index of spatial autocorrelation. *PLoS ONE*, 8(7): e68336

Chen YG (2014). The spatial meaning of Pareto's scaling exponent of city-size distributions. *Fractals*, 22(1-2):1450001

Chen YG (2015a). Power-law distributions based on exponential distributions: Latent scaling, spurious Zipf's law, and fractal rabbits. *Fractals*, 23(2): 1550009

Chen YG (2015b). The distance-decay function of geographical gravity model: power law or exponential law? *Chaos, Solitons & Fractals*, 77: 174-189

Chen YG (2018). Logistic models of fractal dimension growth of urban morphology. *Fractals*, 26(3): 1850033

Chen YG, Feng J (2017). A hierarchical allometric scaling analysis of Chinese cities: 1991-2014. *Discrete Dynamics in Nature and Society*, Volume 2017, Article ID 5243287

Chen YG, Wang YH, Li XJ (2019). Fractal dimensions derived from spatial allometric scaling of urban form. *Chaos, Solitons & Fractals*, 126: 122-134

Clark C (1951). Urban population densities. *Journal of Royal Statistical Society*, 114(4): 490-496

Clauset A, Shalizi CR, Newman MEJ (2009). Power-law distributions in empirical data. *Siam Review*, 51(4): 661-703

Diebold FX (2007). *Elements of Forecasting* (4th edition). Mason, Ohio: Thomson/South-Western

Feder J (1988). *Fractals*. New York: Plenum Press

Frankhauser P (1994). *La Fractalité des Structures Urbaines (The Fractal Aspects of Urban Structures)*. Paris: Economica (In French)

Frankhauser P (1998). The fractal approach: A new tool for the spatial analysis of urban agglomerations. *Population: An English Selection*, 10(1): 205-240

Gao PC, Liu Z, Liu G, Zhao HR, Xie XX (2017). Unified metrics for characterizing the fractal nature of geographic features. *Annals of the American Association of Geographers*, 107(6): 1315-1331

Goldenfeld N, Kadanoff LP (1999). Simple lessons from complexity. *Science*, 284(2): 87-89

Gordon K (2005). The mysteries of mass. *Scientific American*, 293(1):40-46/48

Gould SJ (1979). An allometric interpretation of species-area curves: The meaning of the coefficient. *The American Naturalist*, 114(3): 335-343

Hack JT (1957). Studies of longitudinal streams profiles in Virginia and Maryland. *U.S. Geological Survey Professional Papers B*, 294: 45-97

Hao BL (1986). Fractals and fractal dimensions. *Science*, 38 (1): 9-17 (In Chinese)

Henry J (2002). *The Scientific Revolution and the Origins of Modern Science (2nd Edition)*. New York: Palgrave

Holland J H (1998). *Emergence: from Chaos to Order*. Cambridge, Massachusetts: Perseus Books

Jiang B (2013). Head/tail breaks: A new classification scheme for data with a heavy-tailed distribution. *The Professional Geographer*, 65 (3), 482–494

Jiang B (2015), Geospatial analysis requires a different way of thinking: The problem of spatial heterogeneity, *GeoJournal*, 80(1), 1–13

Jiang B (2015). Head/tail breaks for visualization of city structure and dynamics. *Cities*, 43, 69–77

Jiang B (2018). Spatial heterogeneity, scale, data character and sustainable transport in the big data era. *ISPRS International Journal of Geo-Information*, 7(5), 167

Jiang B, Jia T (2011). Zipf's law for all the natural cities in the United States: a geospatial perspective. *International Journal of Geographical Information Science*, 25(8): 1269-1281

Jiang B, Liu XT (2012). Scaling of geographic space from the perspective of city and field blocks and using volunteered geographic information. *International Journal of Geographical Information Science*, 26(2): 215-229

Jiang B, Yao X (2010, Eds.). *Geospatial Analysis and Modeling of Urban Structure and Dynamics*. New York: Springer

Jiang B, Yin J (2014). Ht-index for quantifying the fractal or scaling structure of geographic features. *Annals of the Association of American Geographers*, 104(3): 530-541

Johnston R (2003). Order in space: geography as a discipline in distance. In: *A Century of British Geography*. Johnston R, Williams M (Eds.). Oxford: Oxford University Press, pp303-345

Kac M (1969). Some mathematical models in science. *Science*, 166: 695-699

Lee Y (1989). An allmetric analysis of the US urban system: 1960–80. *Environment and Planning A*, 21(4): 463–476

Liu SD, Liu SK (1993). *An Introduction to Fractals and Fractal Dimension*. Beijing: China Meteorological Press (In Chinese)

Lobo J, Bettencourt LMA, Strumsky D, West GB (2013). Urban scaling and the production function for cities. *PLoS ONE*, 8(3): e58407

Longley PA (1999). Computer simulation and modeling of urban structure and development. In: *Applied Geography: Principles and Practice*. Ed. M. Pacione. London and New York: Routledge, pp605-

619

Louf R, Barthelemy M (2014a). How congestion shapes cities: from mobility patterns to scaling. *Scientific Reports*, 4: 5561

Louf R, Barthelemy M (2014b). Scaling: lost in the smog. *Environment and Planning B: Planning and Design*, 41: 767-769

Mandelbrot BB (1982). *The Fractal Geometry of Nature.* New York: W. H. Freeman and Company

Mandelbrot BB (1989). Fractal geometry: what is it, and what does it do? *Proceedings of the Royal Society of London A: Mathematical and Physical Sciences*, 423 (1864): 3-16

Neumann J von (1961). Collected works (Vol.6). New York/Oxford: Pergamon Press, 492

Nordbeck S (1971). Urban allometric growth. *Geografiska Annaler B*, 53(1): 54-67

Pumain D, Paulus F, Vacchiani-Marcuzzo C, Lobo J (2006). An evolutionary theory for interpreting urban scaling laws. *Cybergeo: European Journal of Geography*, 343, 20 pages

Smeed RJ (1963). Road development in urban area. *Journal of the Institution of Highway Engineers*, 10(1): 5-30

Stanley HE, Meakin P (1988). Multifractal phenomena in physics and chemistry. *Nature*, 335: 405–409

Takayasu H (1990). *Fractals in the Physical Sciences*. Manchester: Manchester University Press

Tobler W (1970). A computer movie simulating urban growth in the Detroit region. *Economic Geography*, 46(2): 234-240

Tobler W (2004). On the first law of geography: A reply. *Annals of the Association of American Geographers*, 94(2): 304-310

Waldrop M (1992). *Complexity: The Emerging of Science at the Edge of Order and Chaos*. NY: Simon and Schuster

Wang FQ, Li HQ (1996). *Fractals--The Artistic Structure of Nature*. Jinan: Shandong Education Press (In Chinese)

Wheeler JA (1983). Review on *The Fractal Geometry of Nature* by Benoit B. Mandelbrot. *American Journal of Physics*, 51(3): 286-287

Wilson EO (1992). *The Diversity of Life*. Cambridge, Mass: Belknap Press of Harvard University Press